# A statistical physics analysis of expenditure in the UK

**Elvis Oltean, Fedor V. Kusmartsev**

e-mail: elvis.oltean@alumni.lboro.ac.uk

**Abstract:** Most papers which explored so far macroeconomic variables took into account income and wealth. Equally important as the previous macroeconomic variables is the expenditure or consumption, which shows the amount of goods and services that a person or a household purchased. Using statistical distributions from Physics, such as Fermi-Dirac and polynomial distributions, we try to fit the data regarding the expenditure distribution divided in deciles of population according to their income (gross and disposable expenditure are taken into account). Using coefficient of determination as theoretical tool in order to assess the degree of success for these distributions, we find that both distributions are really robust in describing the expenditure distribution, regardless the data set or the methodology used to calculate the expenditure values for the deciles of income. This is the first paper to our knowledge which tackles expenditure, especially using a method to describe expenditure such as lower limit on expenditure. This is also relevant since it allows the approach of macroeconomic systems using more variables characterizing their activity, can help in the investigation of living standards and inequality, and points to more theoretical explorations which can be very useful for the Economics and business practice.

## 1. Introduction

The macroeconomic variables which most papers approach are income and wealth. However, as important as these two variables is expenditure (consumption). Income figures may not always be relevant and sometimes these figures can be quite misleading. Expenditure is a more relevant and concrete variable, since it shows the amount of goods and services that people purchase, which is more relevant for the exploration of living standards and inequality. Expenditure or consumption is an important macroeconomic variable which provides important information about the behavior of the population and helps forecasting the phases of economic cycle (boom and recession). Expenditure or consumption is an important macroeconomic variable which describes the behavior of people companies and/or states.

This is the first paper tackling systematically the expenditure distribution. However, there are several limitations regarding the data made available. The only country which to the best of our knowledge made available such data divided in deciles of population is the UK. We stated this having in mind that a pool of several countries providing similar data would give a better picture of the phenomenon. However, the data provided are very interesting from several points of view. First, the expenditure data that we use in our analysis are about several types of expenditure such as gross expenditure, disposable expenditure, and a specialized type of expenditure, mean disposable expenditure for clothing and footwear. Thus, we plan to use data sets which describe largely the expenditure in the most general forms (namely disposable and gross expenditure) and a narrow category of expenditure. We chose this particular type of expenditure in order to prove the high degree of robustness. Second, the data provided are calculated using two different methodologies such as mean expenditure and lower bound of expenditure decile.

The statistical distributions we use are Fermi-Dirac and polynomial distributions. Both distributions were used successfully in describing income and wealth. While Fermi-Dirac is to some extent famous for being able to analyze the distribution of physical particles such as fermions, polynomials are to a lesser extent known for their applicability to dynamic systems.

## 2. Short Literature Review and Theoretical Framework

Most papers so far used Boltzmann-Gibbs, Bose-Einstein,

and lognormal (Gibrat) distributions in the analysis of some macroeconomic variables such as income and wealth. Mostly, in the statistical analyses authors used data divided in deciles of population ranked according to their increasing values for income or wealth, calculated for a time interval of one year. Also, the most used method/value for deciles was mean (average) value calculated for a part of population contained in a certain decile. Moreover, the most often considered values for the macroeconomic variables were about disposable values, other types of income such as gross income being almost entirely unconsidered [1-6].

More recently, new trends in the analysis of macroeconomic variables using statistical Physics distribution emerged. Thus, [7] and [8] used new types of distributions such as Fermi-Dirac and polynomial distribution. Also, new methods used for calculation of different values for the deciles of data were taken into account. Thus, apart from mean income and mean wealth, upper limit on income/wealth was a methodology used to calculate the income and wealth by using the highest value from the ones ranked increasingly in a decile. The term was used to the best of our knowledge by the national statistical body from Finland [9].

First paper to consider expenditure distribution in passing is [10], which assessed the expenditure distribution in Uganda for three non-consecutive years using Fermi-Dirac and polynomial distributions. For those data, the distributions used proved to be a success given the high values obtained for the coefficient of determination.

## 3. Methodology

The previous paper which tackled the expenditure distribution contained data only from Uganda, which is an underdeveloped country. The shortcoming of the data was about the reliability, since this category of countries has a huge share of black and grey market. In the current paper, given that the provided data are from the UK and that they span over a time interval of 13 years, we consider them to be more reliable in the exploration of the applicability of these distributions. It is noteworthy that some data made available were not valid for calendar year but for time intervals which span partially over two consecutive years [11]. The data are expressed for households in weekly values regarding the expenditure.

Expenditure is calculated based on income deciles. Thus, there are two methodologies used for calculation. For example, mean expenditure is the sum of the total expenditure of the people falling into a certain income decile, divided to the number of people whose income fall into the income range specific to a certain income decile. Regarding the lower bound of expenditure deciles, we have in mind the methodology for upper limit on income, this term being coined by the statistical national body of Finland [10]. Similarly, we use the term lower limit on expenditure in order to designate the lower bound value for expenditure deciles.

In the calculation of the probability, we use cumulative distribution function which expresses the probability (part of the population) which has a certain level of expenditure level above a certain threshold that we chose. We assume that for no expenditure (nil value) the probability (the population) that has higher level is 100%. This assumption is made given that every person needs to spend (at least) partially the personal income in order to survive and cover living expenses. Also, no person is assumed without any income. Even though the ranking is made according to the income values, in general the higher is the income the higher is the level of expenditure. So, the higher is the cardinal number designating an income decile, the higher is the amount of expenditure for that decile.

More formally, the methodology described for mean expenditure data is as follows: let $x_1, x_2, \ldots\ldots\ldots\ldots x_{10}$ be the values assigned for each decile of mean expenditure. Thus, x=0 is represented in the cumulative distribution function with a probability of 100%. Similarly, all the other values are assigned decreasing values for probabilities. Let M be the set of plots of the cumulative distribution function representing the probability for mean expenditure to be expressed graphically. So, M={(0,100%), $(x_1,90\%),(x_2,80\%),(x_3,70\%)$, $(x_4,60\%)$, $(x_5,50\%)$, $(x_6,40\%)$, $(x_7,30\%)$, $(x_8,20\%)$, $(x_9, 10\%)$, $(x_{10}, 0\%)$}.

In the case of lower limit on expenditure, for $x_1=0$ the corresponding probability value is 100% since all expenditure values are higher. Subsequently, all expenditure values are assigned as follows: L={(0,100%), $(x_2,90\%)$, $(x_3,80\%)$, $(x_4,70\%)$, $(x_5,60\%)$, $(x_6,50\%)$, $(x_7,40\%)$, $(x_8,30\%)$, $(x_9, 20\%)$, $(x_{10}, 10\%)$}. In the case of the tenth decile, there are 10% of the population which have a higher expenditure values.

In order to fit the data, we used two distributions. First distribution is the first degree polynomial

$$Y = P_1 * X + P_2 \qquad (1)$$

We considered the first degree polynomial to be the optimal choice regarding the degree/number of coefficients, since the distributions yielded values for coefficient of determination higher than 90%. $P_1$ and $P_2$ are obtained from fitting the data using the first degree polynomial.

The other distribution we used is the Fermi-Dirac distribution [12]. The formula we use is

$$N(\epsilon_i) = \frac{g_i}{exp(\frac{\epsilon_i - \mu}{T}) + 1} \qquad (2)$$

We use this formula in order to have the total number of fermions, which implies the average number over single-particle states. Therefore, the average number of fermions per unit is energy range per unit volume multiplied by degeneracy. The previous relevant literature analogies made between thermodynamics and Macroeconomics use money quantity as an economic analogue for energy ( ) from physical systems. The main parameters used for analysis are degeneracy (c), temperature (T), and chemical potential (μ). The values for Fermi-Dirac distribution are displayed using logarithmic values (log-log scale), unlike the polynomial distribution which uses normal values.

## 4. Results

We present the results obtained from fitting the data using Fermi-Dirac and polynomial distributions in the Tables 1-10. Thus, in the Tables 1-4 are exhibited the coefficients obtained from fitting the data using polynomial applied to general gross and disposable expenditure distribution regardless the methodology to calculate expenditure value for each income decile (i.e. lower limit or mean). Similarly, in the Tables 5-8 we display the results for coefficients from fitting the data using Fermi-Dirac distribution. In the Tables 9-10, we exhibit the results for mean disposable expenditure for clothing and footwear (using both Fermi-Dirac and polynomial distributions). In order to assess the robustness of the distributions, we used coefficient of determination ($R^2$) expressed in percentages.

*Table 1. Coefficients obtained from fitting mean disposable expenditure.*

| Year | P1 | P2 | $R^2$(%) |
|---|---|---|---|
| 2000/2001 | -0.02537 | 85.62 | 93.87 |
| 2001/2002 | -0.02465 | 85.72 | 93.85 |
| 2002/2003 | -0.02414 | 85.83 | 94.00 |
| 2003/2004 | -0.0234 | 85.85 | 94.17 |
| 2004/2005 | -0.02258 | 86.18 | 94.37 |
| 2005/2006 | -0.02222 | 85.93 | 93.95 |
| 2006 | -0.02153 | 86.37 | 94.47 |
| 2007 | -0.02145 | 86.44 | 94.41 |
| 2008 | -0.02083 | 85.7 | 93.83 |
| 2009 | -0.0218 | 87.22 | 94.90 |
| 2010 | -0.02088 | 86.93 | 94.62 |
| 2011 | -0.02038 | 86.9 | 94.83 |
| 2012 | -0.02023 | 87.04 | 94.89 |

*Table 2. Coefficients obtained from fitting mean gross expenditure.*

| Year | P1 | P2 | $R^2$(%) |
|---|---|---|---|
| 2000/2001 | -0.02536 | 85.51 | 93.89 |
| 2001/2002 | -0.02461 | 85.54 | 93.83 |
| 2002/2003 | -0.02409 | 85.68 | 94.00 |
| 2003/2004 | -0.02336 | 85.72 | 94.12 |
| 2004/2005 | -0.02251 | 86 | 94.38 |
| 2005/2006 | -0.02251 | 86 | 94.38 |
| 2006 | -0.02153 | 86.28 | 94.46 |
| 2007 | -0.02141 | 86.39 | 94.48 |
| 2008 | -0.02079 | 85.63 | 93.95 |
| 2009 | -0.02174 | 86.92 | 94.87 |
| 2010 | -0.02086 | 86.85 | 94.70 |
| 2011 | -0.02035 | 86.84 | 94.88 |
| 2012 | -0.02023 | 87.01 | 94.90 |

*Table 3. Coefficients obtained from fitting lower limit on disposable expenditure.*

| Year | P1 | P2 | $R^2$(%) |
|---|---|---|---|
| 2000/2001 | -0.02593 | 86.21 | 93.22 |
| 2001/2002 | -0.02405 | 86.01 | 93.00 |
| 2002/2003 | -0.02325 | 86.42 | 93.39 |
| 2003/2004 | -0.0231 | 86.61 | 93.64 |
| 2004/2005 | -0.0218 | 86.62 | 93.57 |
| 2005/2006 | -0.02131 | 86.29 | 93.15 |
| 2006 | -0.02046 | 86.38 | 93.26 |
| 2007 | -0.01981 | 86.6 | 93.48 |
| 2008 | -0.01574 | 84.62 | 91.46 |
| 2009 | -0.01923 | 86.65 | 93.44 |
| 2010 | -0.01891 | 86.49 | 93.25 |
| 2011 | -0.01814 | 86.95 | 93.75 |
| 2012 | -0.01819 | 86.98 | 93.80 |

*Table 4. Coefficients obtained from fitting lower limit on gross expenditure.*

| Year | P1 | P2 | $R^2$(%) |
|---|---|---|---|
| 2000/2001 | -0.0214 | 84.84 | 91.83 |
| 2001/2002 | -0.01987 | 84.61 | 91.53 |
| 2002/2003 | -0.01936 | 85.05 | 91.99 |
| 2003/2004 | -0.01904 | 85.14 | 92.16 |
| 2004/2005 | -0.01792 | 85.1 | 92.04 |
| 2005/2006 | -0.01752 | 84.8 | 91.61 |
| 2006 | -0.01682 | 84.92 | 91.77 |
| 2007 | -0.01619 | 85.04 | 91.94 |
| 2008 | -0.01574 | 84.62 | 91.46 |
| 2009 | -0.01601 | 85.2 | 91.96 |
| 2010 | -0.01581 | 85.1 | 91.8 |
| 2011 | -0.01517 | 85.49 | 92.25 |
| 2012 | -0.01537 | 85.58 | 92.34 |

*Table 5. Coefficients obtained from fitting mean disposable expenditure.*

| Year | T | C | μ | $R^2$(%) |
|---|---|---|---|---|
| 2000/2001 | 0.4999 | 4.4 | 8.113 | 98.61 |
| 2001/2002 | 0.4974 | 4.401 | 8.139 | 98.60 |
| 2002/2003 | 0.4952 | 4.4 | 8.164 | 98.61 |
| 2003/2004 | 0.4881 | 4.393 | 8.196 | 98.50 |
| 2004/2005 | 0.4866 | 4.398 | 8.235 | 98.60 |
| 2005/2006 | 0.4931 | 4.404 | 8.241 | 98.71 |
| 2006 | 0.4887 | 4.404 | 8.285 | 98.80 |
| 2007 | 0.4837 | 4.4 | 8.287 | 98.52 |
| 2008 | 0.4999 | 4.402 | 8.309 | 98.56 |
| 2009 | 0.4662 | 4.4 | 8.273 | 98.60 |
| 2010 | 0.4763 | 4.402 | 8.316 | 98.70 |
| 2011 | 0.4748 | 4.4 | 8.343 | 98.67 |
| 2012 | 0.4654 | 4.396 | 8.347 | 98.55 |

*Table 6. Coefficients obtained from fitting mean gross expenditure.*

| Year | T | C | μ | $R^2$(%) |
|---|---|---|---|---|
| 2000/2001 | 0.4965 | 4.396 | 8.11 | 98.55 |
| 2001/2002 | 0.4992 | 4.4 | 8.139 | 98.57 |
| 2002/2003 | 0.495 | 4.398 | 8.164 | 98.61 |
| 2003/2004 | 0.4919 | 4.394 | 8.198 | 98.47 |
| 2004/2005 | 0.4852 | 4.393 | 8.237 | 98.53 |
| 2005/2006 | 0.4751 | 4.401 | 8.063 | 98.66 |
| 2006 | 0.486 | 4.401 | 8.282 | 98.72 |
| 2007 | 0.4797 | 4.396 | 8.288 | 98.56 |
| 2008 | 0.4952 | 4.397 | 8.31 | 98.51 |
| 2009 | 0.4671 | 4.398 | 8.272 | 98.61 |
| 2010 | 0.4719 | 4.398 | 8.315 | 98.63 |
| 2011 | 0.4713 | 4.397 | 8.343 | 98.66 |
| 2012 | 0.4634 | 4.395 | 8.344 | 98.55 |

**Table 7.** Coefficients obtained from fitting lower limit on disposable expenditure.

| Year | T | C | μ | $R^2$(%) |
|---|---|---|---|---|
| 2000/2001 | 0.5616 | 4.41 | 8.218 | 98.87 |
| 2001/2002 | 0.5677 | 4.41 | 8.294 | 98.89 |
| 2002/2003 | 0.5576 | 4.41 | 8.328 | 98.87 |
| 2003/2004 | 0.5495 | 4.408 | 8.334 | 98.83 |
| 2004/2005 | 0.5532 | 4.411 | 8.393 | 98.91 |
| 2005/2006 | 0.5648 | 4.413 | 8.416 | 98.89 |
| 2006 | 0.5614 | 4.412 | 8.457 | 98.90 |
| 2007 | 0.5555 | 4.412 | 8.489 | 98.89 |
| 2008 | 0.6104 | 4.414 | 8.717 | 98.88 |
| 2009 | 0.5575 | 4.415 | 8.519 | 98.97 |
| 2010 | 0.5632 | 4.416 | 8.536 | 98.96 |
| 2011 | 0.5493 | 4.414 | 8.578 | 98.94 |
| 2012 | 0.5474 | 4.413 | 8.575 | 98.94 |

**Table 8.** Coefficients obtained from fitting lower limit on gross expenditure.

| Year | T | C | μ | $R^2$(%) |
|---|---|---|---|---|
| 2000/2001 | 0.5977 | 4.41 | 8.409 | 98.88 |
| 2001/2002 | 0.607 | 4.412 | 8.483 | 98.88 |
| 2002/2003 | 0.5952 | 4.412 | 8.51 | 98.87 |
| 2003/2004 | 0.5896 | 4.409 | 8.527 | 98.84 |
| 2004/2005 | 0.594 | 4.413 | 8.587 | 98.93 |
| 2005/2006 | 0.6066 | 4.416 | 8.611 | 98.91 |
| 2006 | 0.6012 | 4.414 | 8.651 | 98.91 |
| 2007 | 0.5963 | 4.413 | 8.689 | 98.88 |
| 2008 | 0.6104 | 4.414 | 8.717 | 98.88 |
| 2009 | 0.5976 | 4.417 | 8.701 | 98.98 |
| 2010 | 0.603 | 4.42 | 8.714 | 98.99 |
| 2011 | 0.5904 | 4.418 | 8.755 | 98.98 |
| 2012 | 0.5905 | 4.418 | 8.756 | 98.98 |

**Table 9.** Coefficients obtained from fitting clothing and footwear mean disposable expenditure.

| Year | T | C | μ | $R^2$(%) |
|---|---|---|---|---|
| 2000/2001 | 0.5318 | 4.402 | 5.264 | 98.68 |
| 2001/2002 | 0.5243 | 4.404 | 5.283 | 98.63 |
| 2002/2003 | 0.5361 | 4.403 | 5.259 | 98.55 |
| 2003/2004 | 0.5353 | 4.402 | 5.267 | 98.41 |
| 2004/2005 | 0.5355 | 4.41 | 5.327 | 98.52 |
| 2005/2006 | 0.5135 | 4.413 | 5.259 | 98.84 |
| 2006 | 0.547 | 4.423 | 5.298 | 99.11 |
| 2007 | 0.5231 | 4.41 | 5.23 | 98.72 |
| 2008 | 0.549 | 4.414 | 5.224 | 98.76 |
| 2009 | 0.5053 | 4.399 | 5.158 | 98.69 |
| 2010 | 0.565 | 4.415 | 5.294 | 98.74 |
| 2011 | 0.505 | 4.411 | 5.225 | 98.93 |
| 2012 | 0.523 | 4.41 | 5.279 | 98.84 |

**Table 10.** Coefficients obtained from fitting clothing and footwear mean disposable expenditure.

| Year | P1 | P2 | $R^2$(%) |
|---|---|---|---|
| 2000/2001 | -0.4387 | 84.52 | 93.11 |
| 2001/2002 | -0.4267 | 84.75 | 92.86 |
| 2002/2003 | -0.435 | 84.12 | 92.59 |
| 2003/2004 | -0.4286 | 83.99 | 92.36 |
| 2004/2005 | -0.4085 | 84.59 | 92.70 |
| 2005/2006 | -0.434 | 85.29 | 93.16 |
| 2006 | -0.4188 | 84.44 | 92.63 |
| 2007 | -0.4452 | 84.69 | 92.91 |
| 2008 | -0.4502 | 84.09 | 92.29 |
| 2009 | -0.4711 | 84.6 | 92.75 |
| 2010 | -0.4146 | 83.29 | 91.49 |
| 2011 | -0.4526 | 85.66 | 93.68 |
| 2012 | -0.4205 | 84.57 | 92.51 |

Using Fermi-Dirac distribution, the annual values for coefficient of determination yielded from fitting the data are higher than 98% for all data sets (both for disposable and gross income and as well for mean and lower limit methods of calculation).

Similarly, using polynomial distribution to fit the annual data, the results for coefficient of determination are generally higher than 93%. There is an exception regarding the lower limit gross income, for which the values regarding coefficient of determination were slightly lower (above 91%). The explanation regarding the slight disparity in the values for the coefficient of determination is attributable to the number of coefficients, as Fermi-Dirac distribution has three parameters, while first degree polynomial distribution has only two parameters.

Regarding the influence of the methodology in calculating the expenditure values for deciles, there are no significant changes for the annual values of the coefficient of determination between lower limit expenditure sets and mean expenditure sets in the case of Fermi-Dirac distribution. In the case of polynomial distribution, lower limit gross expenditure set yielded values for coefficient of determination which are slightly lower (91%) compared to the mean counterpart values. Most of the values for coefficient of determination regarding mean gross expenditure are around 93 %.

As for mean disposable expenditure for clothing and footwear data set, the values for coefficient of determination are around 91-93 % for polynomial distribution and 98 % for Fermi-Dirac distribution. These values are in accordance to the general values for expenditure obtained for general types of expenditure (gross and disposable expenditure).

In the Figures 1 and 2, we display few typical graphical results. The Figure 1 is about Fermi-Dirac distribution, while the Figure 2 is about polynomial distribution. The values for coefficient of determination are 98.47% for the graph displayed in Figure 1 and 91.46% for the graph displayed in Figure 2.

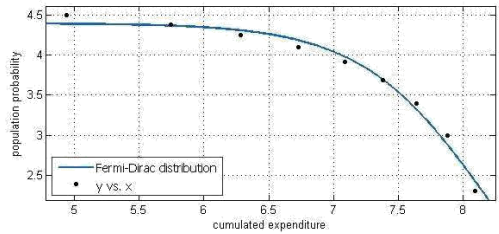

*Figure 1. Fermi-Dirac distribution applied to mean gross expenditure from the year 2003/2004.*

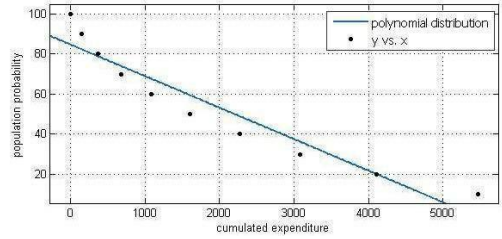

*Figure 2. Polynomial distribution applied to lower limit on gross expenditure from the year 2008.*

## 5. Conclusions

The expenditure (consumption) can be described successfully both by Fermi-Dirac and polynomial distributions. Thus, these distributions are robust in describing the phenomenon, given that the annual values for coefficient of determination are higher than 91% in all cases. Also, the robustness is maintained regardless the data sets (gross or disposable) or methodology for calculating the values for expenditure deciles (mean or lower limit values) that we used. Moreover, both Fermi-Dirac and polynomial distributions are more powerful in describing expenditure distribution than Bose-Einstein distribution especially considering the values for coefficient of determination.

The results from fitting the data can be dramatically improved in the case of the polynomial distribution if a higher degree polynomial is used (at least a second degree polynomial). We chose to use first degree polynomial for our analysis in order to have minimal number of parameters.

From the point of view of Economic theory, the expenditure distribution can help in finding possible causes for ups and downs in the economic cycle considering that demand influences it to a large extent. Also, this may help in clarifying the relation which exists between income and expenditure, which was first described by Keynes. Moreover, having information simultaneously about income and expenditure can help clarifying the evolution of savings in a macroeconomic system.

The expenditure distribution may help to clarify the impact of the taxation system and benefits at macroeconomic level, given that gross and disposable expenditure data were provided. This is important for any economic system, considering that taxation system can increase or impact negatively the overall macroeconomic evolution. Also, benefits system can help in shedding light on the welfare state theory and practice.

For business theory and practice, the expenditure distribution can help in the marketing policies of the companies in order to forecast the demand for certain targeted groups. Thus, it is very useful for a company to know the upcoming evolution of the expenditure in order to estimate the future sales.